\documentclass[aps,twocolumn,superscriptaddress,pra]{revtex4-1}

\usepackage{pdfpages} 
\makeatletter
\AtBeginDocument{\let\LS@rot\@undefined}
\makeatother

\usepackage{amsfonts}
\usepackage{amsmath}
\usepackage{amssymb}
\usepackage{bbold}
\usepackage[colorlinks,
	linkcolor=red,
	citecolor=blue,
	urlcolor=red]{hyperref}
\usepackage{graphicx}
\usepackage[all]{hypcap}

\newcommand{\figref}[1]{Fig.~\ref{#1}}
\newcommand{\equref}[1]{eq.~(\ref{#1})}

\newcommand{\mec}{\mathrm{m}}
\newcommand{\cav}{\mathrm{c}}
\newcommand{\aux}{\mathrm{aux}}
\newcommand{\eff}{\mathrm{eff}}
\newcommand{\eigenfreq}{\tilde{\omega}}
\newcommand{\modea}{\hat a}
\newcommand{\modeb}{\hat b}

\begin{document}

\title{Level attraction in a microwave optomechanical circuit}

\author{N.~R.~Bernier}
\affiliation{Institute of Physics, \'Ecole Polytechnique F\'ed\'erale de Lausanne (EPFL), CH-1015 Lausanne, Switzerland}
\author{L.~D.~T\'{o}th}
\affiliation{Institute of Physics, \'Ecole Polytechnique F\'ed\'erale de Lausanne (EPFL), CH-1015 Lausanne, Switzerland}
\author{A.~K.~Feofanov}
\email[]{alexey.feofanov@epfl.ch}
\affiliation{Institute of Physics, \'Ecole Polytechnique F\'ed\'erale de Lausanne (EPFL), CH-1015 Lausanne, Switzerland}
\author{T.~J.~Kippenberg}
\email[]{tobias.kippenberg@epfl.ch}
\affiliation{Institute of Physics, \'Ecole Polytechnique F\'ed\'erale de Lausanne (EPFL), CH-1015 Lausanne, Switzerland}

\begin{abstract}
  Level repulsion --
  the opening of a gap between two degenerate modes due to coupling
  --
  is ubiquitous anywhere
  from solid state theory to quantum chemistry. 
  In contrast, 
  if one mode has negative energy,
  the mode frequencies attract instead.
  They converge and develop imaginary components,
  leading to an instability;
  an exceptional point marks the transition.
  This, however, only occurs if
  the dissipation rates of the two modes are comparable.
  Here we 
  expose a theoretical framework for the general phenomenon
  and realize it experimentally
  through engineered dissipation
  in a multimode superconducting microwave optomechanical circuit.
  Level attraction is observed 
  for a mechanical oscillator and a superconducting microwave cavity,
  while an auxiliary cavity is used for sideband cooling.
  Two exceptional points are demonstrated that
  could be exploited for their topological properties.
\end{abstract}
\maketitle

Level repulsion of two coupled modes 
with an energy crossing 
has applications ranging from
solid state theory~%
\cite{ashcroft_solid_1976}
to quantum chemistry~%
\cite{atkins_atkins_2009}.
While deceptively simple, 
it spawns a wealth of physics.
With the introduction of dissipation or gain,
an exceptional point~%
\cite{heiss_physics_2012}
appears
that is topologically non-trivial~%
\cite{dembowski_experimental_2001,uzdin_observability_2011,milburn_general_2015}.
The special case of two modes with equal dissipation and gain rates 
is an example of parity-time symmetry~%
\cite{bender_introduction_2005,bender_observation_2013}.
The spontaneous breaking of that symmetry
is marked by the exceptional point.
In recent years, 
exceptional points gathered significant interest 
and they were demonstrated
in a variety of systems including
active microwave circuits~%
\cite{stehmann_observation_2004,schindler_experimental_2011,assawaworrarit_robust_2017},
lasers~%
\cite{brandstetter_reversing_2014,peng_loss-induced_2014}
and optical microresonators~%
\cite{bender_twofold_2013,peng_parity-time-symmetric_2014,chen_exceptional_2017}.
In particular, the topological transfer of energy 
between states by circling
an exceptional point has been demonstrated
with a microwave cavity~%
\cite{dembowski_experimental_2001},
a microwave waveguide~%
\cite{doppler_dynamically_2016},
as well as an optomechanical system~%
\cite{xu_topological_2016,jing_PT-symmetric_2014}.

Strikingly, 
if one mode has negative energy,
the energy levels of two interacting modes 
do not repel,
but attract instead~%
\cite{bernier_unstable_2014,eleuch_exceptional_2014,seyranian_coupling_2005}.
The  Hamiltonian leads to hybridized modes
of complex eigenfrequencies, one of which is unstable.
As in level repulsion, an exceptional point marks the transition
between the regimes of real and complex frequencies.
In the process, the real components of the frequencies become identical
in a way that is reminiscent of the synchronization of driven oscillators~%
\cite{pikovsky_synchronization:_2003}.

\begin{figure}[ht!]
  \begin{center}
      \includegraphics{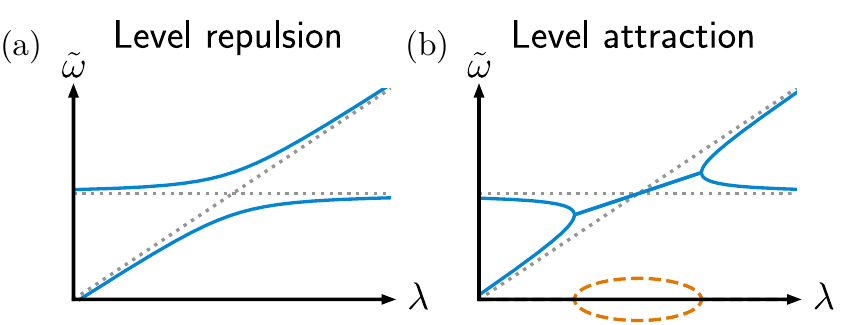}
  \end{center}
  \caption{
  Level repulsion and attraction.
  Two modes,
  whose bare frequencies  depend on a parameter $\lambda$,
  have a level crossing (dotted lines).
  A coherent coupling will in general lift the degeneracy.
  (a)~In the more usual case, level repulsion,
  the coupling opens a gap between the frequencies 
  of the hybridized eigenmodes $\eigenfreq$ (blue solid lines)
  and the eigenfrequencies bend away from each other.
  (b)~In contrast, if one of the modes
  has negative energy, 
  level attraction occurs.
  The real components
  of the eigenfrequencies $\eigenfreq$ (blue solid lines)
  bend towards each other and converge.
  They meet at two exceptional points,
  where the curves have kinks.
  A gap opens in the imaginary components of the frequencies
  (orange dashed lines).
  The mode with a negative
  imaginary component to the frequency
  is unstable and grows exponentially.
  }
  \label{fig:LAvsLR}
\end{figure}

Negative-energy modes
(equivalent to harmonic oscillators with negative mass)
have been studied
in schemes to evade quantum measurement backaction~%
\cite{tsang_evading_2012,zhang_back-action-free_2013,polzik_trajectories_2015}.
Such a scheme was recently demonstrated
with an atomic spin ensemble,
prepared in its maximal-energy spin state
in a magnetic field~%
\cite{moller_quantum_2017}.
Spin flips decrease the energy and
correspond to excitations of
a  harmonic oscillator with a negative mass.
Alternatively, 
the negative-energy mode can be effectively realized 
in a frame rotating faster than the mode itself~%
\cite{woolley_two-mode_2013,ockeloen-korppi_quantum_2016}.

In cavity optomechanics~%
\cite{aspelmeyer2014}, 
a blue-detuned pump tone
induces time-dependent interactions between
the electromagnetic mode and the mechanical oscillator.
In a frame rotating at the pump frequency,
the Hamiltonian is time-independent,
and 
the electromagnetic mode appears to have negative energy. 
While level repulsion was demonstrated
in the strong coupling regime of cavity optomechanics~%
\cite{teufel_circuit_2011,verhagen_quantum-coherent_2012},
level attraction has so far not been observed.

Here 
we construct a general theoretical framework to understand
the phenomenon
and, as an illustration,
demonstrate level attraction in 
a microwave optomechanical circuit using engineered dissipation.
In a first part, 
it is shown how a coherent coupling
between modes of positive and negative energy
gives rise to level attraction.
The role of dissipation is discussed
and explains the difficulty in observing level attraction
in such systems, as the dissipation rates of the two modes 
must be similar.
An intuitive way to classify different types of exceptional points
in two-modes system is developed that allows to clearly distinguish
the cases of level repulsion and attraction.
In a second part, both level attraction and repulsion 
are demonstrated experimentally in the same 
microwave optomechanical circuit,
where the mechanical dissipation rate 
can be engineered to match that of the microwave cavity.


We start with a general theoretical model of
a positive-energy mode coherently coupled to a negative-energy one.
The two modes, of annihilation operators
$\modea$ and $\modeb$ and 
coherently coupled with strength $g$,
are described by the Hamiltonian
\begin{equation}
  \hat H =
  - \hbar \omega_1 (\lambda) \modea^\dagger \modea
  +
  \hbar \omega_2 (\lambda) \modeb^\dagger \modeb
  +
  \hbar g \left( \modea\modeb + \modea^\dagger \modeb^\dagger \right)
  \label{eq:hamiltLA}
\end{equation}
where the two positive frequencies
$\omega_1$ and $\omega_2$
vary with respect to an external parameter 
$\lambda$.
The linear coupling chosen here is quite general:
if we assume the modes close in frequency, 
other linear terms 
$\modea^\dagger \modeb$, $\modea\modeb^\dagger$
can be neglected in the rotating wave approximation
(valid only if the frequencies $\omega_{1,2}$
dominate over the dissipation rates for an open system).
The coupling rate
$g$
is chosen to be real, as any complex phase can be absorbed in
a redefinition of
$\modea$ or $\modeb$.

In the Heisenberg picture, this leads to the equations of motion
\begin{equation}
  \frac{d}{dt}
  \begin{pmatrix}
    \modea \\ \modeb^\dagger
  \end{pmatrix}
  =
  i
  \begin{pmatrix}
    \omega_1 & -g \\
    g & \omega_2
  \end{pmatrix}
  \begin{pmatrix}
    \modea \\ \modeb^\dagger
  \end{pmatrix}
  \label{eq:EOM}
\end{equation}
where we drop the explicit 
$\lambda$
dependence.
We note that the uncoupled, bare modes evolve as
$\modea(t) = e^{i\omega_1 t} \modea(0)$
and
$\modeb^\dagger(t) = e^{i\omega_2 t} \modeb^\dagger(0)$
with a positive phase.
The hybridized eigenmodes of the system are found 
by diagonalizing the matrix in \equref{eq:EOM},
and have eigenfrequencies
\begin{equation}
  \eigenfreq_{1,2}
  =
  \frac{\omega_1 + \omega_2}{2}
  \pm
  \sqrt{
  \left( \frac{\omega_1 - \omega_2}{2} \right)^2
  - g^2
  }.
  \label{eq:eigenfreqs}
\end{equation}
The negative sign in front of 
$g^2$
is the only difference with the eigenfrequencies for the case 
of level repulsion (when $\modea$ has positive energy)
but dramatically impacts on the physics.

In \figref{fig:LAvsLR},
level attraction is compared to level repulsion,
with two striking features.
First, instead of avoiding each other, 
the eigenfrequencies pull towards each other.
Second, when they meet at
$4g^2 = (\omega_1 - \omega_2)^2$,
the frequencies
acquire positive and negative imaginary parts,
causing exponential decay and growth.
The hybridized mode with a negative imaginary component 
grows exponentially and is therefore unstable.

\begin{figure}[]
  \begin{center}
      \includegraphics{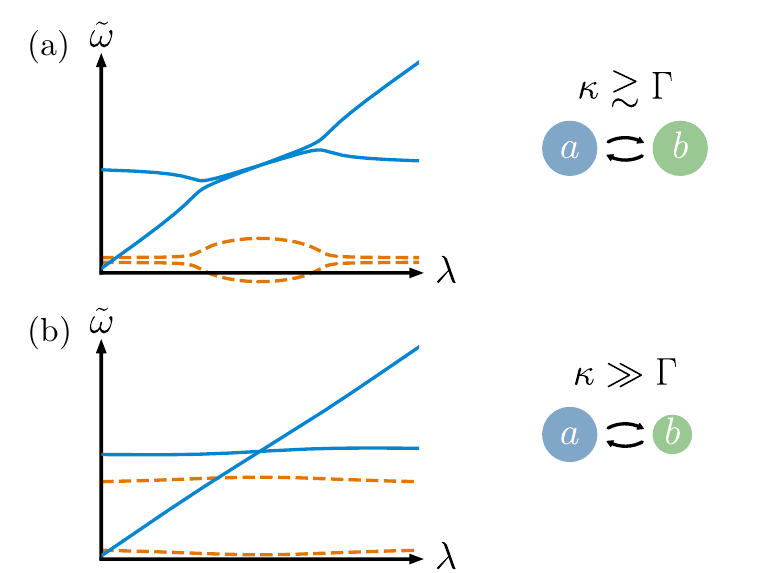}
  \end{center}
  \caption{
  The effect of dissipation on level attraction.
  Two modes, $\modea$ and $\modeb$ cross in frequency
  (the mode $\modea$ having negative energy),
  with respective dissipation rates
  $\kappa$ and $\Gamma$.
  The real component of the frequency is the solid blue line
  and the imaginary component the dashed orange line.
  (a)~While a finite average dissipation rate simply translates
  the imaginary components of the frequency, 
  a difference in the two rates ($\kappa\neq\Gamma$)
  affects qualitatively level attraction.
  The kinks and the exceptional points disappear
  and the picture is overall smoothed out.
  (b)~When the dissipation rates differ significantly,
  no trace of level attraction is visible anymore.
  In both cases, one of the hybridized mode becomes unstable 
  if the imaginary component of its frequency turns negative.
  }
  \label{fig:dissipation}
\end{figure}

The transition between the regimes of real and complex
eigenfrequencies is marked by exceptional points,
which can be understood by studying  the 
matrix of \equref{eq:EOM}.
Decomposed in terms of Pauli matrices
and omitting the term proportional to the identity,
it can be expressed as
$
\frac{1}{2} (\omega_1 - \omega_2) \sigma_z
- ig \sigma_y
$.
In contrast with level repulsion
for which the interaction term would be
$g\sigma_x$,
here the Hermitian Pauli matrix is multiplied by
an \emph{imaginary} coefficient.
The transition between the two regimes corresponds to
a competition of the two terms.
When the two Pauli matrices have coefficients 
of the same amplitude, the matrix is proportional to
$\sigma_z - i\sigma_y$.
At this point, the two eigenvectors coalesce and 
a single eigenvector with a single eigenvalue subsists:
it is an exceptional point~%
\cite{heiss_physics_2012}.
More generally for all two-mode systems, 
any point when the dynamics is determined by
a matrix proportional to 
$\sigma_\alpha + i \sigma_\beta$,
with  
$\alpha\neq \beta$,
is an exceptional point.
In the supplementary information,
we use this decomposition to construct 
an intuitive classification of the various realizations of
exceptional points.

Level attraction arises whenever
the coupling term consists of a Pauli matrix 
with an imaginary coefficient.
In fact, coupled oscillators of positive and negative energy
are only one way to achieve this.
An alternative relies on
dissipative interaction
between two modes
through one or multiple intermediary modes~%
\cite{metelmann_quantum-limited_2014}.
The mode hybridization observed between
positive-energy oscillators
with dissipative interactions~%
\cite{xu_topological_2016,%
gloppe_bidimensional_2014,khanbekyan_unconventional_2015}
can be interpreted as level attraction.

While level attraction of two linearly coupled modes
displays intriguing similarities with the synchronization
of driven oscillators, important differences exist.
As in synchronization,
the real components of the frequencies ``lock'' over a frequency
range that increases with the coupling rate $g$,
and form the equivalent of an Arnold tongue~%
\cite{pikovsky_synchronization:_2003}.
The physical process however differ.
In synchronization, 
one starts with two oscillators 
that are driven nonlinearly
to their limit-cycles,
then a coupling is introduced that locks their frequencies 
and their phases~%
\cite{weiss_noise-induced_2016}.
In level attraction by contrast,
the frequencies of the two modes attract
through \emph{linear} dynamics until they become identical.
The state of the two hybridized modes remain independent
and their phases can be set arbitrarily.

To understand why level attraction is in practice 
less common than level repulsion, 
the role of dissipation
should be studied.
We open the system and include in our treatment 
the energy dissipation rates
$\kappa$ and $\Gamma$ respectively
for the modes $\modea$ and $\modeb$.
They can be introduced as positive imaginary components 
of the bare frequencies in the equations of motion.
The results of \equref{eq:EOM} and (\ref{eq:eigenfreqs}) 
can be extended by replacing
$\omega_1$ with
$\omega_1 + i \frac{\kappa}{2}$
and $\omega_2$ with
$\omega_2 + i \frac{\Gamma}{2}$.
In \figref{fig:dissipation}, we compare
the resulting eigenfrequencies.
If the dissipation rates are equal ($\kappa=\Gamma$),
the level structure of \figref{fig:LAvsLR}b is reproduced
with the imaginary components translated 
to a finite average.
However, in the case of even slightly mismatched dissipation rates
$\kappa\gtrsim\Gamma$ (\figref{fig:dissipation}a),
the exceptional points and the kinks in the frequencies
all disappear.
For increasingly dissimilar rates 
$\kappa \gg \Gamma$ (\figref{fig:dissipation}b),
the level-attraction picture progressively disappears
until the modes seem to cross without interacting.
Therefore, only in a system where dissipation rates can be tuned
to closely match each other 
is level attraction observable.

Cavity optomechanics provides an ideal setting to study
level attraction and compare it to level repulsion.
We now take $\modea$ to represent an electromagnetic mode
and $\modeb$ a mechanical oscillator, coupled through
the optomechanical interaction
$\hbar g_0 \modea^\dagger \modea (\modeb + \modeb^\dagger)$,
where 
$g_0$
is the vacuum optomechanical coupling~%
\cite{aspelmeyer2014}.
With a blue-detuned pump tone applied to the system,
the three-wave-mixing coupling is linearized
and the Hamiltonian reduces to 
the form of \equref{eq:hamiltLA}
\begin{equation}
  \hat H = 
  - \hbar \Delta \modea^\dagger \modea
  +
  \hbar \Omega_\mec \modeb^\dagger \modeb
  + \hbar g
  \left( \modea \modeb + \modea^\dagger \modeb^\dagger \right),
  \label{eq:OMHamilt}
\end{equation}
where
$\Delta$
is the detuning of the pump tone,
$\Omega_\mec$
the mechanical mode frequency and
$g = g_0 \sqrt{n_\cav}$
the linear coupling enhanced by the mean cavity photon number
$n_\cav$
due to the pump tone.
As above, 
we neglect counter-rotating terms
and assume the detuning 
$\Delta$
to be close to
$\Omega_\mec$.
Critically, the Hamiltonian is expressed in a frame 
rotating at the pump frequency in order to be time-independent.
Hence, for a blue detuning
$\Delta>0$,
the cavity mode effectively has a \emph{negative} energy,
since  
the photons have a negative
relative frequency with respect to the pump.
In this context, 
the well-known parametric instability of optomechanics~%
\cite{aspelmeyer2014}
can be interpreted as resulting from the physics of level attraction. 
The instability stems from 
the negative imaginary component that develops
in the eigenfrequencies of the equations of motion,  
above the critical coupling
$g_\mathrm{crit} = \sqrt{\kappa \Gamma}/2$.
For level attraction to be observable,
the magnitudes of $\kappa$ and $\Gamma$ should be close.
For usual experimental parameters, however, 
the electromagnetic decay rate $\kappa$
is much larger than the mechanical rate $\Gamma$,
and no attraction can be observed in practice
for the mechanical and electromagnetic modes.

\begin{figure}[]
  \begin{center}
      \includegraphics{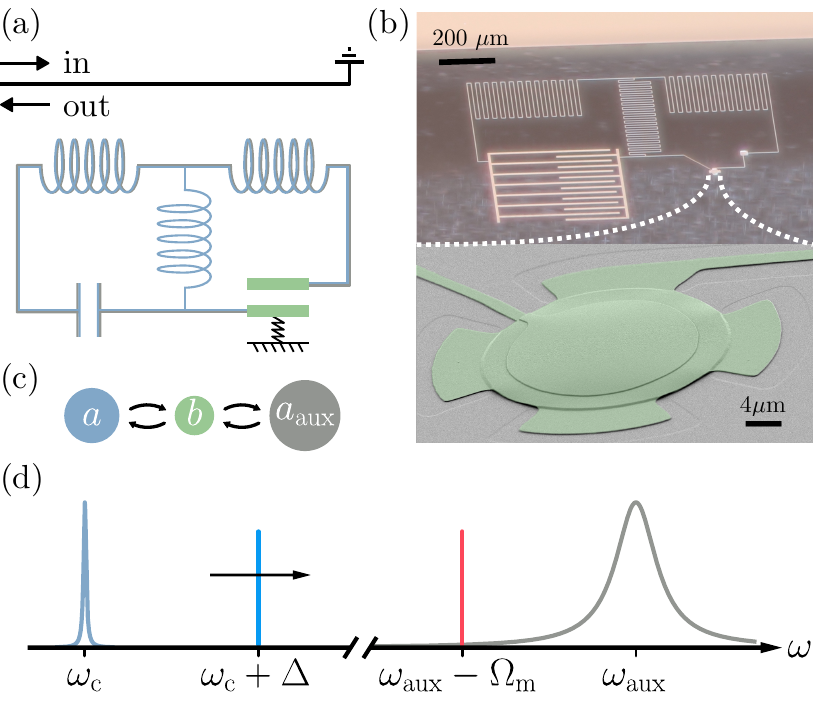}
  \end{center}
  \caption{
  Engineering dissipation in a multimode optomechanical circuit.
  In order to observe level attraction, the dissipation rate
  $\Gamma$ of the mechanical mode $\modeb$
  must be increased to match $\kappa$, 
  the much larger dissipation rate of the
  primary electromagnetic mode $\modea$.
  To that end, an auxiliary mode $\modea_\aux$
  is used for sideband cooling.
  (a)~Schematic of the microwave optomechanical circuit,
  coupled inductively to a 
  microwave feedline and measured in reflection.
  The two hybridized modes of the circuit 
  $\modea$ and $\modea_\aux$ interact
  with the motion of the top membrane
  of a shared capacitor,
  acting as the mechanical oscillator $\modeb$ (in green).
  (b)~Photograph of the circuit and 
  scanning-electron micrograph of the 
  vacuum-gap capacitor.
  (c)~Diagram of the three interacting modes.
  (d)~Frequency domain representation of the level-attraction experiment.
  A microwave pump tone (vertical red line), 
  red-detuned by the mechanical frequency 
  $\Omega_\mec$
  with respect to the auxiliary mode resonance $\omega_\aux$ 
  (grey peak) is used for sideband cooling.
  Level attraction of the modes $\modea$ and $\modeb$
  is achieved by sweeping the detuning $\Delta$
  of a pump tone (vertical blue line) 
  near the blue sideband of 
  the primary mode resonance $\omega_\cav$ (blue peak).
  For level repulsion,
  the pump tone is instead swept near the red sideband.
  }
  \label{fig:scheme}
\end{figure}

\begin{figure*}[t]
  \begin{center}
      \includegraphics{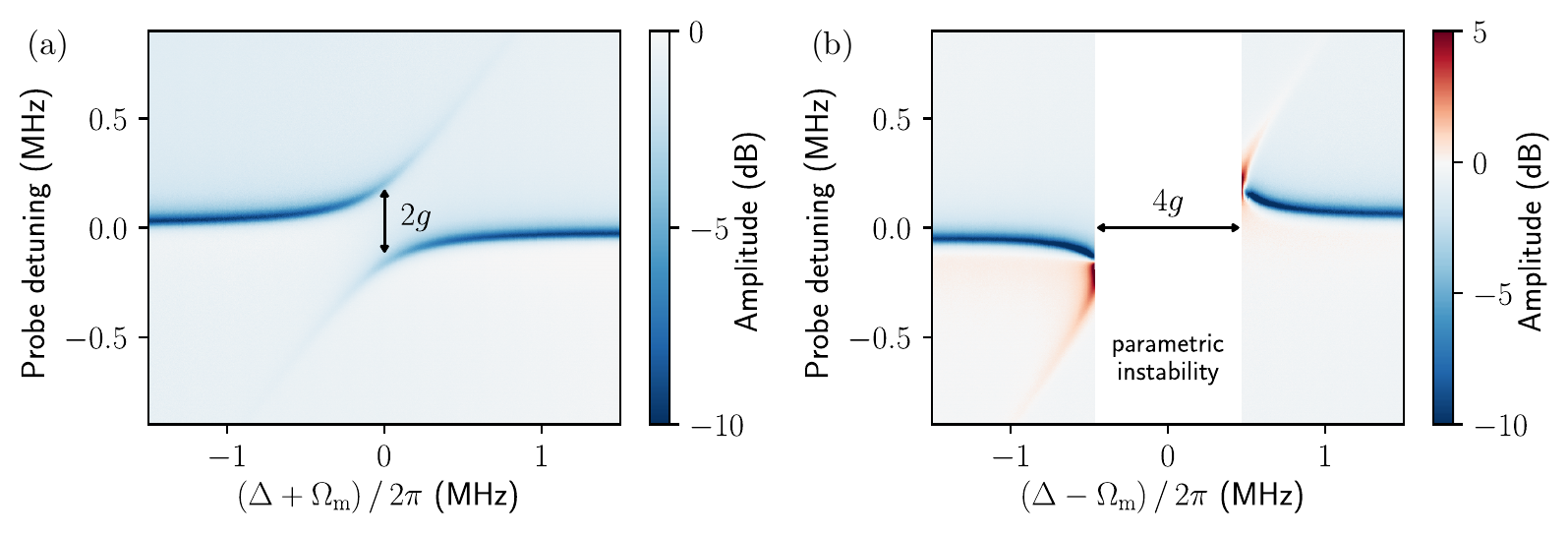}
  \end{center}
  \caption{
  Experimental demonstration of level repulsion and attraction
  in a microwave optomechanical circuit.
  Amplitude response of the system 
  as the detuning $\Delta$ of the pump tone is varied,
  when the effective dissipation rate
  of the mechanical mode $\modeb$ matches that 
  of the microwave mode $\modea$.
  In the laboratory frame, $\Delta$
  determines the effective frequency of the mechanical oscillator,
  that is swept across the microwave resonance.
  (a)~When the pump tone is swept in frequency 
  across the red sideband of the microwave mode, 
  the two resonances bend away from each other.
  (b)~If the pump tone is swept across the blue sideband instead,
  the resonances pull towards each other and converge
  near two exceptional points.
  Data is omitted for clarity
  where the system becomes unstable
  and one hybridized mode grows exponentially 
  until the conditions of validity of \equref{eq:OMHamilt} 
  are no longer fulfilled (parametric instability).
  }
  \label{fig:LAresult}
\end{figure*}

In our experiment,
the effective mechanical energy decay rate 
$\Gamma_\eff$
is artificially increased 
to match $\kappa$
using sideband cooling with an auxiliary mode.
We use a superconducting electromechanical circuit~%
\cite{teufel_circuit_2011}
containing two microwave LC modes
interacting with 
the vibrational mode
of a vacuum-gap capacitor
(represented schematically in 
\figref{fig:scheme}a and shown in \figref{fig:scheme}b).
The design, which was
demonstrated in previous work~%
\cite{toth_dissipative_2017},
uses two hybridized electromagnetic modes of the circuit
to ensure that one has
a much larger external coupling rate
to the microwave feedline
than the other.
The more dissipative, auxiliary mode $\modea_\aux$ 
is used to perform sideband cooling
of the mechanical oscillator with a red-detuned pump tone.
This damps the oscillator and increases its effective
dissipation rate to
$\Gamma_\eff\approx\kappa$.
Meanwhile, the less dissipative, primary mode $\modea$
undergoes level attraction 
with the damped mechanical oscillator.

In the experiment, the device
is placed inside a dilution refrigerator and cooled 
to the base temperature below 50~mK,
at which
the circuit is superconducting
and therefore its internal $Q$-factor is enhanced.
The two microwave modes $\modea$ and $\modea_\aux$
have respective resonance frequencies 
$\omega_\cav \approx 2\pi\times 4.11$~GHz and 
$\omega_\aux \approx 2\pi\times 5.22$~GHz,
and dissipation rates
$\kappa \approx 2\pi\times 110$~kHz and
$\kappa_\aux \approx 2\pi \times 1.8$~MHz.
They interact with the fundamental mode of the top plate
of the vacuum-gap capacitor 
that has a frequency $\Omega_\mec \approx 2\pi \times6.3$~MHz.
By placing a pump tone red-detuned by $\Omega_\mec$ 
from the auxiliary mode resonance 
(see \figref{fig:scheme}b),
the mechanical oscillator is damped.
The mechanical dissipation rate $\Gamma$,
originally below $ 2\pi \times 100$~Hz, 
is tuned to 
an effective dissipation rate 
$\Gamma_\eff \approx \kappa \approx 2\pi \times 110$~kHz.

Level repulsion and attraction of 
the primary microwave mode and the damped mechanical oscillator
are both measured.
As a pump tone 
is tuned to the blue (or red) sideband of 
the primary microwave mode
(\figref{fig:scheme}b),
the weak probe tone of a vector network analyser 
is applied to obtain its linear response.
Due to the hybridization of the modes, the response carries information
about both microwave and mechanical modes.
In both cases, the same pump power 
is set to obtain a coupling strength
$g \approx 2\pi \times 200$~kHz
corresponding to a mean cavity photon number
$n_\cav \approx 4\times 10^6$.
The known case of level repulsion is obtained with a red-detuned tone
(\figref{fig:LAresult}a).
As the bare effective mechanical mode frequency 
comes near the microwave resonance,
the two modes hybridize; 
their eigenfrequencies bend away from each other
and a gap of $2g$ opens.
If a blue-detuned tone is used instead, 
level attraction occurs, 
shown in \figref{fig:LAresult}b
which displays the characteristic level structure of
\figref{fig:LAvsLR}b.
The resonance frequencies of the modes attract and converge 
at the points where
the bare frequencies of the modes differ only by  
$\pm2g$.
Data is omitted for clarity in the unstable region where
the real component of the frequencies are identical.
In order for the level attraction to be clearly visible,
a large coupling rate $g$ 
is chosen that exceeds the dissipation rates 
$\kappa$ and $\Gamma_\eff$.
It  therefore exceeds the critical coupling $g_\mathrm{crit}$ as well
and parametric instability occurs:  
in the unstable region,
one of the modes grows exponentially 
until the conditions of the validity 
of \equref{eq:OMHamilt} are no longer fulfilled.
Namely, the fluctuating field is no more negligible
compared to the mean cavity photon number $n_\cav$.
The original nonlinear optomechanical interaction 
constrain the system to a limit-cycle with 
a modified cavity resonance frequency, the description of which
lies beyond the scope of this article~%
\cite{marquardt_dynamical_2006}.

In summary, level attraction was experimentally demonstrated 
using a dual-mode electromechanical circuit.
Although related to the well-studied parametric instability
of optomechanics,
the vastly different dissipation rates for the 
mechanical and electromagnetic modes prevented its observation until now.
Level attraction, 
similarly to level repulsion in open systems,
gives rise to exceptional points.
In both cases,
the real part of the frequencies converge
and a gap opens in the imaginary part (or vice versa)
precisely at the exceptional point.
In future work, the exceptional points of level attraction
could be harnessed to demonstrate topological phenomena
by circling such a point in
a two-dimensional parameter space~%
\cite{dembowski_experimental_2001,xu_topological_2016,doppler_dynamically_2016}.
Since the exceptional point only exists when the dissipation rates of
the two modes match exactly,
the tunable mechanical damping rate 
$\Gamma_\eff$
can be used as one parameter
in such an experiment, 
with the tunable coupling rate
$g$
as the second.

\begin{acknowledgments}
  We thank A.~Nunnenkamp and D.~Malz for useful discussions
  and comments.
  This work was supported by the SNF, 
  the NCCR Quantum Science and Technology (QSIT),  
  and the EU Horizon 2020 research
  and innovation programme under grant agreement No 732894
  (FET Proactive HOT). 
  TJK acknowledges financial support from an ERC AdG (QuREM).
  All samples were fabricated in the Center of Micro\-Nano\-Technology (CMi) at EPFL.
\end{acknowledgments}

\newpage


\section*{Supplementary material}

# Supplementary Information - Level attraction in a microwave optomechanical circuit

N. R. Bernier,[1] L. D. Tóth,[1] A. K. Feofanov,[1] and T. J. Kippenberg[1]

[1]*Institute of Physics, École Polytechnique Fédérale de Lausanne (EPFL), CH-1015 Lausanne, Switzerland*

## I. SYMMETRY BETWEEN LEVEL REPULSION AND ATTRACTION

We explicitly derive here two minimal models, one for the usual level repulsion of two coherently coupled modes, the other for level attraction of a negative-energy mode coherently coupled to a positive-energy mode. We show that a symmetry relation links the two cases and all the physics of one is mirrored in the other if the frequency and dissipation rates are exchanged.

First, we review level repulsion. Two modes of positive energy (with annihilation operators $\hat{a}$ and $\hat{b}$, and respective frequencies $\omega_1$ and $\omega_2$) interact, as described by the Hamiltonian

$$\hat{H}_{\mathrm{LR}} = \hbar\omega_1 \hat{a}^\dagger \hat{a} + \hbar\omega_2 \hat{b}^\dagger \hat{b} + \hbar g \left( \hat{a}\hat{b}^\dagger + \hat{a}^\dagger \hat{b} \right) \tag{1}$$

with $g$ the linear coupling strength. The equation of motion in the Heisenberg picture is given by

$$\frac{d}{dt}\begin{pmatrix} \hat{a} \\ \hat{b} \end{pmatrix} = -i \begin{pmatrix} \omega_1 & g \\ g & \omega_2 \end{pmatrix} \begin{pmatrix} \hat{a} \\ \hat{b} \end{pmatrix}. \tag{2}$$

To solve the system in terms of eigenmodes, the matrix is diagonalized. The eigenfrequencies are given by

$$\tilde{\omega}_{1,2}^{\mathrm{LR}} = \frac{\omega_1 + \omega_2}{2} \pm \sqrt{\left(\frac{\omega_1 - \omega_2}{2}\right)^2 + g^2}. \tag{3}$$

The model is extended to describe modes with dissipation by adding imaginary components to the bare frequencies, substituting $\omega_1 \to \omega_1 - i\Gamma_1/2$ and $\omega_2 \to \omega_2 - i\Gamma_2/2$. Note that a negative sign is required to obtain decaying exponentials. For simplicity, we define $\omega_0 = \frac{\omega_1+\omega_2}{2}$, $\Gamma_0 = \frac{\Gamma_1+\Gamma_2}{2}$ and $\Delta\omega = \omega_1 - \omega_2$, $\Delta\Gamma = \Gamma_1 - \Gamma_2$. The eigenfrequencies can now be written as

$$\tilde{\omega}_{1,2}^{\mathrm{LR}} = \omega_0 - i\frac{\Gamma_0}{2} \pm \sqrt{\left(\frac{\Delta\omega - i\Delta\Gamma/2}{2}\right)^2 + g^2}. \tag{4}$$

The frequency of oscillation of the eigenmodes is given by the real component of $\tilde{\omega}_{1,2}^{\mathrm{LR}}$ and the dissipation rate by half its imaginary component (with a minus sign).

We now consider the case of level attraction. One (and only one) mode has negative energy. The system is now described by the Hamiltonian

$$\hat{H}_{\mathrm{LA}} = -\hbar\omega_1 \hat{a}^\dagger \hat{a} + \hbar\omega_2 \hat{b}^\dagger \hat{b} + \hbar g \left( \hat{a}\hat{b} + \hat{a}^\dagger \hat{b}^\dagger \right). \tag{5}$$

The equations of motion are given by

$$\begin{pmatrix} \hat{a} \\ \hat{b}^\dagger \end{pmatrix} = i \begin{pmatrix} \omega_1 & -g \\ g & \omega_2 \end{pmatrix} \begin{pmatrix} \hat{a} \\ \hat{b}^\dagger \end{pmatrix} \tag{6}$$

with eigenfrequencies

$$\tilde{\omega}_{1,2}^{\mathrm{LA}} = \frac{\omega_1 + \omega_2}{2} \pm \sqrt{\left(\frac{\omega_1 - \omega_2}{2}\right)^2 - g^2}. \tag{7}$$

The only difference with eq. (3) is the sign in front of $g^2$, which can result in a complex eigenvalue meaning an instability for the system. It might appear at first disconcerting that the Hermitian Hamiltonian in eq. (5) leads to complex eigenfrequencies and unstable dynamics. In fact, for such an infinite Hilbert space, an operator must be compact as well as Hermitian to guarantee the existence of real eigenvalues [1], which is not the case here. We also

note that only when the eigenfrequencies are real can the eigenoperators be interpreted as Bogolyubov modes [2]. When the eigenfrequencies are complex, the required commutation relations cannot be satisfied.

In order to include dissipation, we now substitute $\omega_1 \to \omega_1 + i\Gamma_1/2$ and $\omega_2 \to \omega_2 + i\Gamma_2/2$. Note the opposite sign as above is required to obtain decaying exponentials. The eigenfrequencies including dissipation become

$$\tilde{\omega}_{1,2}^{\text{LA}} = \omega_0 + i\Gamma_0/2 \pm \sqrt{\left(\frac{\Delta\omega + i\Delta\Gamma/2}{2}\right)^2 - g^2}. \tag{8}$$

There is a symmetry between eq. (4) and eq. (8). They are equivalent under the transformation $\omega' = \Gamma/2$ and $\Gamma' = 2\omega$, if eq. (8) is multiplied by a factor $-i$:

$$\tilde{\omega}_{1,2}^{\prime\text{LA}} = \omega_0' - i\Gamma_0'/2 \pm \sqrt{\left(\frac{\Delta\omega' - i\Delta\Gamma'/2}{2}\right)^2 + g^2}. \tag{9}$$

We conclude that level attraction and repulsion are equivalent to each other through the exchange of frequency and dissipation rates (within a factor of 2). In Fig. S1, this symmetry is highlighted by contrasting equivalent situations for both level repulsion and level attraction Hamiltonians. In Fig. S1a-d, the curves for the real and imaginary parts of the eigenfrequencies are interchanged when going from the left column (level repulsion) to the right (level attraction). Moreover, Fig. S1e-f can be compared with Fig. 1 of the main text, where the characteristic shape of level attraction is here seen for the dissipation rates for the Hamiltonian of level repulsion and vice versa. In general, for $\hat{H}_{\text{LR}}$ the *frequencies repel*, while the *dissipation rates attract*. The exact opposite is true for $\hat{H}_{\text{LA}}$: it is the *dissipation rates* that *repel* and the *frequencies* that *attract*.

## II. CLASSIFICATION OF EXCEPTIONAL POINTS FOR SYSTEMS OF TWO COUPLED MODES

We describe here how to classify exceptional points of $2 \times 2$ matrices using Pauli matrices and proceed to sort out under which cases fall recent experimental demonstrations of exceptional points.

In general, the equations of motion for a 2-modes system can be written in the form

$$i\frac{d}{dt}\begin{pmatrix}\hat{d}_1\\ \hat{d}_2\end{pmatrix} = M\begin{pmatrix}\hat{d}_1\\ \hat{d}_2\end{pmatrix} \tag{10}$$

where $\hat{d}_1$, $\hat{d}_2$ are the operators for the two modes and $M$ a $2 \times 2$ matrix. The eigenmodes of the system and their eigenfrequencies correspond to the eigenvectors and eigenvalues of $M$. If for some parameter values, $M$ only has a single eigenvector and a single eigenvalue, this is called an exceptional point. The matrix $M$ can always be decomposed in terms of the Pauli matrices, defined as

$$\sigma_x = \begin{pmatrix}0 & 1\\ 1 & 0\end{pmatrix}, \ \sigma_y = \begin{pmatrix}0 & -i\\ i & 0\end{pmatrix}, \text{ and } \sigma_z = \begin{pmatrix}1 & 0\\ 0 & -1\end{pmatrix}. \tag{11}$$

in the form

$$M = a_0 \mathbb{1} + a_1 \sigma_x + a_2 \sigma_y + a_3 \sigma_z \tag{12}$$

with $\mathbb{1}$ the identity matrix and $a_i$ complex numbers. The eigenvalues of $M$ are now easily expressed as

$$\lambda_{1,2} = a_0 \pm \sqrt{a_1^2 + a_2^2 + a_3^2}. \tag{13}$$

Note that the first term proportional to the identity in eq. (12) only shifts the eigenvalues by a constant and has no effect on the eigenvectors. Anytime that the matrix $M$ can be written as a multiple of $\sigma_\alpha + i\sigma_\beta$ ($\alpha \neq \beta$) (plus the identity), there is only one eigenvalue and this is an exceptional point [3]. We can use this decomposition to classify examples of exceptional points.

(I) The most common case is the level repulsion of two (positive-energy) modes of degenerate frequencies due to a coherent coupling. Most of the experimentally demonstrated exceptional points have realized such a system [4–11]. The two modes are $\hat{d}_1 = \hat{a}$ and $\hat{d}_2 = \hat{b}$ in our previous notation. The matrix $M$ can be written as

$$M = \left(\omega_0 - i\frac{\Gamma_0}{2}\right)\mathbb{1} - i\frac{\Delta\Gamma}{4}\sigma_z + g\sigma_x. \tag{14}$$

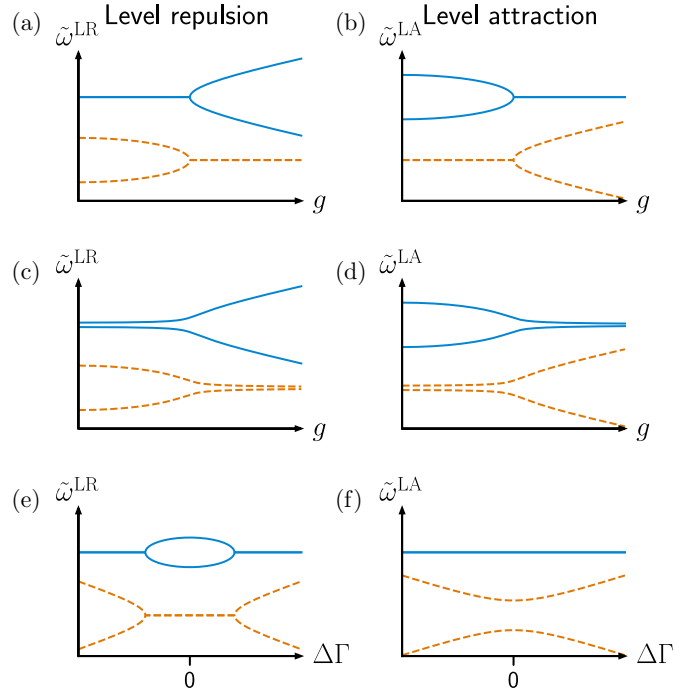

FIG. S1. Comparison of the real (solid blue lines) and imaginary (dashed orange lines) components of the eigenfrequencies $\tilde{\omega}$ for the level-repulsion Hamiltonian $\hat{H}_{\mathrm{LR}}$ (left column) and the level-attraction Hamiltonian $\hat{H}_{\mathrm{LA}}$ (right column). (a,b) The transition through an exceptional point, as a function of the coupling strength $g$. For level repulsion (a), the modes are originally degenerate ($\Delta\omega = 0$), while for level attraction (b) they originally have matching dissipation rates ($\Delta\Gamma = 0$). (c,d) Same as (a,b) with a slight degeneracy ($\Delta\omega \neq 0$ for level repulsion (c) and $\Delta\Gamma \neq 0$ for level attraction (d)), such that the exceptional point is avoided. (e,f) The difference in dissipation rates $\Delta\Gamma$ is varied, for degenerate modes $\Delta\omega = 0$. Note that this is the complementary case to Fig. 1 of the main text, where $\Delta\Gamma = 0$ and $\Delta\omega$ is varied. For level repulsion (e), the imaginary component of the frequency attract and converge in a region where a gap opens in the real components of the frequency (analogously to level attraction as a function of $\Delta\omega$). For level attraction (f), the real component is unchanged while the dissipation rates have an avoided crossing (analogously to level repulsion as a function of $\Delta\omega$).

For low coupling rate $g$, $M$ is diagonal with a gap in dissipation rates. For large $g$, the last term dominates such that the eigenmodes are eigenvectors of $\sigma_x$: $g$ opens a gap in frequency, while the dissipation rates are shared equally. The transition between the regimes is marked by an exceptional point at $g = \Delta\Gamma/4$.

(II) In this article, we consider the case of level attraction of modes of negative and positive energies and matching dissipation rates. Here the operators are $\hat{d}_1 = \hat{a}$ and $\hat{d}_2 = \hat{b}^\dagger$ in our notation. The matrix $M$ can be written as

$$-M = \left(\omega_0 + i\frac{\Gamma_0}{2}\right)\mathbb{1} + \frac{\Delta\omega}{2}\sigma_z - ig\sigma_y. \tag{15}$$

At low coupling $g$, there is a gap in frequency, while at high coupling, the $\sigma_y$ term opens a gap in dissipation rates and the frequencies are identical. An exceptional point marks the transition at $g = \Delta\omega/2$. Note that the coherent coupling corresponds to a term with an imaginary coefficient for a system with one mode of negative energy.

(III) Level attraction of two modes can be realized in any system in which the coupling term has an imaginary component. In particular, dissipative interactions [12] represent an alternative way to the one presented in this article. The matrix $M$ is there expressed as

$$M = \left(\omega_0 - i\frac{\Gamma_{\mathrm{eff}}}{2}\right)\mathbb{1} + \frac{\Delta\omega}{2}\sigma_z + ig_{\mathrm{dis}}\sigma_x \tag{16}$$

where for simplification the effective dissipation rate $\Gamma_{\mathrm{eff}}$ of the two coupled modes is taken to be approximately equal. The effective interaction between the two modes has an imaginary coefficient and they have an increased

effective dissipation rate $\Gamma_{\text{eff}}$ due to the dissipative interaction as well. Similarly to the previous case, by increasing the coupling $g_{\text{dis}}$ the original gap in frequency is closed and a difference in dissipation rates is created. In contrast however, $\Gamma_{\text{eff}}$ grows proportionally with $g_{\text{dis}}$, such that the gap in dissipation rates does not result in an instability. The mode hybridization between modes of positive energy coupled with dissipative interactions can be interpreted as level attraction. This is the case for the experiments by Xu *et al.* [13], where two mechanical oscillators are effectively coupled by both interacting with the same optical cavity, by Gloppe *et al.* [14] where two modes of a nanowire interact through the non-conservative force of an optical field, and by Khanbeykyan *et al.* [15] where two modes of an optical resonator interact through multiple quantum dots.

(IV) Finally, yet another way to implement an exceptional point was realized in the experiment of Chen *et al.* [16]. The clockwise and counterclockwise modes of a whispering-gallery-mode resonator (of degenerate frequencies and dissipation rates) interact through two Rayleigh scatterers. This results in a combination of coherent and dissipative interaction that can be described by

$$M = \left(\omega_0 - i\frac{\Gamma_0}{2}\right)\mathbb{1} + g_{\text{coh}}\sigma_x + ig_{\text{dis}}\sigma_y. \tag{17}$$

As the coupling $g_{\text{coh}}$ and $g_{\text{dis}}$ are varied, the relative phases of the bare modes change in the hybrid eigenmodes. When the two coupling strengths match ($g_{\text{coh}} = g_{\text{dis}}$), the two eigenmodes coalesce and there is an exceptional point. Interestingly, this corresponds to a point of maximal nonreciprocity, with one bare mode coupled to the second but not the other way around [17].



\end{document}